\documentclass[twocolumn,nofootinbib,prl,aps,showpacs,
tightenlines]{revtex4}
\usepackage{color}
\usepackage{bm}
\usepackage{epsfig}
\def\lsim{\hbox{ \raise.35ex\rlap{$<$}\lower.6ex\hbox{$\sim$}\ }}
\def\gsim{\hbox{ \raise.35ex\rlap{$>$}\lower.6ex\hbox{$\sim$}\ }}
\def\la{\langle}\def\ra{\rangle}
\begin{document}
\title{Kaon-Soliton Bound State Approach to the Pentaquark States}
\author{Byung-Yoon Park$^a$}
\email{bypark@cnu.ac.kr}
\author{Mannque Rho$^b$}
\email{rho@spht.saclay.cea.fr}
\author{Dong-Pil Min$^c$}
\email{dpmin@phya.snu.ac.kr} \affiliation{$^a$ CSSM, University of
Adelaide, Adelaide 5005, Australia
and \\
Department of Physics,Chungnam National University, Daejon 305-764, Korea \\
$^b$ Service de Physique Th\'eorique, CEA Saclay,
91191 Gif-sur-Yvette, France \\
and
Department of Physics, Hanyang University, Seoul 133-791, Korea \\
$^c$ School of Physics and Center for Theoretical Physics,
Seoul National University, Seoul 151-747, Korea}
\date{\today}
\begin{abstract}
We show that in hidden local symmetry theory with the vector
manifestation (VM), a $K^+$ can be bound to skyrmion to give an
exotic baryon that has the quantum numbers of the $\Theta^+$
pentaquark with spin 1/2 and even parity which is consistent with
large $N_c$ counting. The vector meson $K^*$ subject to the VM in
the chiral limit plays an essential role in inducing the binding.
\end{abstract}

\maketitle

The discovery of a $\Theta^+$ baryon\cite{Experiments} with a mass
of 1540 MeV and with a narrow width less than 25 MeV is one of the
most exciting events in recent hadron physics. It is an
interesting exotic state that cannot be simply made of three
quarks as the other baryons. The existence of such an exotic state
has been anticipated also in various models.\cite{Speculations}
Among them, the most precise predictions for the mass and the
width were made in the SU(3) chiral soliton model by Diakonov,
Petrov and Polyakov\cite{DPP97}. There, it is postulated that the
$N(1710)$ and $\Sigma(1880)$ states are the members of the
antidecuplets and then the mass formula coming from the SU(3)
rigid rotator quantization of the Skyrme model is used to predict
the other members such as the isosinglet $\Theta^+$ and the
isoquartet $\Xi$'s in this multiplet. The predicted mass and width
of $\Theta^+$ are surprisingly close to the experimental data.

On the other hand, the large strange quark mass compared with the
two non-strange light-quark masses renders the SU(3) rigid rotor
quantization problematic and in particular for the exotic state,
it has been argued~\cite{cargese-igor,Cohen03,Jenkins04} that its
excitation energy of order $N_c^0$ is inconsistent with the scale
separation needed to justify collective coordinate quantization.
Since the Skyrme model is considered to be a valid description of
the baryons in the large-number-of-colors ($N_c$) limit, such an
$N_c$ inconsistency could be a severe defect although the results
are satisfactory phenomenologically. In Ref.\cite{Cohen03}, it has
even been argued that the collective exotic states cannot be the
genuine prediction at large $N_c$. This argument was given a
further support by an exactly solvable model~\cite{cohen2}.

The objective of this paper is to show that the bound-state
approach first proposed by Callan and
Klebanov\cite{CK,cargese-igor} for the non-exotic $S=-1$ hyperons,
in which the fluctuations in the strangeness direction are of
order $N_c^0$ with a vibrational character and those in the light
degrees of freedom are of order $1/N_c$ with a rotational
character, can be suitably applied to hidden local symmetry
Lagrangian~\cite{HLS,HY:PR} to obtain a bound or quasi-bound
$K^+$-soliton having the quantum numbers of the exotic
pentaquarks.

In fact, Itzhaki {\it et al.\/}\cite{IKOR03,klebanov2} recently
studied this $K^+$-soliton system using the usual three-flavor
Skyrme Lagrangian consisting of the octet pseudo-Goldstone fields
and arrived at the conclusion that with the parameters appropriate
for the $S=-1$ hyperons, the repulsive WZ term for the $K^+$
channel prevented the binding and there can at best be a
near-threshold $S=+1$ bound state -- which can be quantized to
$I=0$, $J^P=1/2^+$ pentaquark state -- {\it only when} the SU(3)
symmetry is strongly broken and the strength of the WZ term is
reduced. However, in contrast to the $S=-1$ case where the bound
state approach and the rigid rotator approach match consistently
to each other with the bound state turning into the rotor
zero-mode, the absence of the $S=+1$ bound state in the SU(3)
symmetric limit raises a serious question on the validity of the
rigid rotor approach where $S=+1$ exotic states come out
independent of the strange quark mass. Our study reaches a similar
conclusion in the chiral limit.

In this work, we show that when vector mesons are incorporated
into the chiral Lagrangian as was done in \cite{SMNR89}, a
dramatic change can take place in the structure of the $S=+1$
skyrmions. In particular, we will see that the hidden local
symmetry (HLS) theory~\cite{HLS} endowed with the vector
manifestation (VM) as formulated recently by Harada and
Yamawaki~\cite{HY:PR} can render under certain conditions that are
not unreasonable the $K^+$ meson bound to an SU(2) skyrmion \`a la
Callan and Klebanov to give a state with the quantum numbers of
the $\Theta^+$. The crucial element in the treatment is that the
vector mesons need to figure explicitly with the VM since as will
be clarified below, the role of the vector mesons becomes more
prominent for the $K^+$-nucleon interactions. There are
qualitatively two imoportant mechanisms at work in arriving at our
result. First, the vector meson $K^*$ exerts a level repulsion
from above, thereby weakening the``pushing-up" effect of the
repulsive WZ term for the $S=+1$ kaon state. Next, the explicit
vector degrees of freedom with masses subject to the
VM~\cite{HY:PR} soften the contact interaction terms between the
pseudoscalars, reducing the strength of the WZ term. That vector
mesons are important for bound pentaquarks for heavy flavors has
been understood since some time~\cite{OPM94}: This is intimately
connected with heavy-quark symmetry. The significant new element
in our theory is that the VM to which HLS theory flows as the
spontaneously broken chiral symmetry is restored renders feasible
a systematic chiral perturbation calculation with the vector
mesons treated on the same footing as the pseudo-Goldstone
bosons~\cite{HY:PR}. A striking case in support of this argument
is the recent analysis of the chiral doublers in light-heavy quark
hadrons starting from the VM fixed point and taking into account
small deviations from the VM in low order perturbation theory
confirms that the VM is not too far from nature~\cite{HRS}.

To bring out the above points in the simplest form, we consider
the HLS Lagrangian for three flavors~\cite{HLS} with only the
relevant degrees of freedom retained. The normal part of the
Lagrangian will be taken in the form
\begin{eqnarray}
{\cal L} &=& \frac{f_\pi^2}{4}
  \left\{
    \mbox{Tr}({\cal D}_\mu \xi^{}_L \xi^\dagger_L
            - {\cal D}_\mu \xi^{}_R \xi^\dagger_R )^2
  \right.
\nonumber\\
&&
%\hskip 2em
+ \mbox{Tr}({\cal M}
 (\xi^\dagger_L \xi^{}_R + \xi^\dagger_R \xi^{}_L -2)
\label{Lag} \\
&&
%\hskip 2em
  \left.
 +a
    \mbox{Tr}({\cal D}_\mu \xi^{}_L \xi^\dagger_L
            + {\cal D}_\mu \xi^{}_R \xi^\dagger_R )^2
  \right\}
%\nonumber\\
%&&
 -\frac{1}{2g^2} \mbox{Tr} F_{\mu\nu} F^{\mu\nu}\nonumber
\nonumber
\end{eqnarray}
where $\xi^\dagger_L \xi_R = U\in SU(3)$ describes the
pseudoscalar octets with masses ${\cal
M}=\mbox{diag}(m_\pi^2,m_\pi^2,m_K^2)$ and the covariant
derivatives are defined as ${\cal D}_\mu = \partial_\mu - i V_\mu$
with the vector meson nonets $V_\mu$ taken as hidden gauge fields
(with the $\omega$ included as the isosinglet vector meson).
$F_{\mu\nu} = {\cal D}_\mu V_\nu - {\cal D}_\nu V_\mu$ is the
field strength tensor. The constant $a$ which will play a crucial
role in our work represents the ratio $(f_\sigma/f_\pi)^2$ where
$f_\sigma$ is the decay constant of the scalar field that gets
``eaten up" to give the vector meson mass and $f_\pi$ the Goldtone
pion decay constant.

We shall take the anomalous part of the Lagrangian which plays a
key role in our treatment in the form
\begin{equation}
{\cal L}_{\mbox{\scriptsize an}}
  = {\cal L}_{\mbox{\scriptsize WZ}}^0
    - 10C ({\cal L}_1 - {\cal L}_2),
\label{Lag_an}\end{equation} where ${\cal L}_{\mbox{\scriptsize
WZ}}^0$ comes from the five-dimensional Wess-Zumino action that we
shall refer to as ``irreducible" WZ term and we have chosen only a
special combination of two terms among four general homogeneous
solutions of the anomaly equation of Ref.\cite{Fujiwara}. Other
choices are found not to affect the essential feature of our
results. Explicitly, they are
${\cal L}_1 = \varepsilon^{\mu\nu\lambda\rho}
 \mbox{Tr}(L_\mu L_\nu L_\lambda R_\rho - R_\mu R_\nu R_\lambda L_\rho)$,
and
${\cal L}_2 = \varepsilon^{\mu\nu\lambda\rho}
\mbox{Tr}(L_\mu R_\nu L_\lambda R_\rho)$
with $L_\mu(R_\mu) = {\cal D}_\mu
\xi^{}_{L(R)} \xi^\dagger_{L(R)}$ and $C=-iN_c/240\pi^2$. This
particular combination of the homogeneous solutions makes the
amplitude of the five-pseudoscalar process $K^+(k^+)+K^-(k^-)
\rightarrow \pi^+(q^+)\pi^-(q^-)\pi^-(q^0)$ entirely given by
$\Gamma^0_{\mbox{\scriptsize WZ}}$ in the chiral limit to be
consistent with QCD anomaly.\cite{SRM88}

The basic premise of the Harada-Yamawaki approach is that the
parameters of the Lagrangian are to be determined \`a la
Harada-Yamawaki~\cite{HY:PR} by matching the effective theory
(via, e.g., correlators) to QCD at a matching scale $\Lambda$ near
the chiral scale $\sim 4\pi f_\pi\sim 1$ GeV. Physical quantities
such as the masses and coupling constants of the fluctuating
fields are obtained by doing loop calculations with the
Lagrangian. This means that not all the parameters that figure for
the soliton justified at the large $N_c$ limit are physical ones.
The quantum corrections are generically suppressed by $1/N_c$
factors. Now what figures the most importantly for our problem is
the parameter $a$ which ranges between 1 and 2 in nature. The
vector meson mass arising from Higgsing in HLS theory is $ m_V^2 =
a g^2 f_\pi^2 $.

At the VM fixed point, $a=1$. The Harada-Yamawaki study shows that
in the large $N_c$ limit, $a$ is $ \sim 1.3$ but at the scale
corresponding to the on-shell vector meson, loop corrections make
$a$ approach 2 at which point the vector meson dominance is
exact~\cite{HY:VDM}. On the other hand, there is a compelling
indication that in the background of baryonic matter, $a$ is close
to 1 with the vector dominance maximally violated~\cite{BR2003}
even though the gauge coupling $g$ departs from the VM fixed-point
value. Following the modern literature~\cite{GV}, we call this
point with $a=1$ and $g\neq 0$ ``Georgi vector limit (GVL)." This
indicates that the range of mass parameters that would figure in
the soliton and in fluctuations in the strangeness direction would
be that closer to 1. It is intriguing that a detailed
analysis~\cite{HY:VDM,HY:PR} indicates that $a=2$ is ``accidental"
and nature is closer to 1 -- the Georgi vector limit (GVL) -- with
the vector dominance violated maximally although other parameters
(such as $f_\pi$) are subject to significant quantum corrections.
It is also intriguing that on a much more fundamental level, $a=1$
represents the ``the theory-space locality" in the little Higgs
mechanism for the $\pi^+$-$\pi^0$ electromagnetic mass
difference~\cite{pimass,GV}. Thus our focus then will be the
property of the soliton for $a$ {\it in the vicinity of 1}.

For comparison with the work of Itzhaki et al~\cite{IKOR03}, we
consider also the limit -- which is artificial in the model --
where $a\rightarrow \infty$ while keeping the KSRF relation
with the parameters $f_\pi$ and $g$ fixed finite.
Then, the second term in the Lagrangian (\ref{Lag}) containing $a$
gives constraints on the vector meson octets as
%\begin{equation}
$V^\infty_\mu = - v_\mu \label{V_octet}$
%\end{equation}
and on the isoscalar $\omega$ as
%\begin{equation}
$\omega^{\infty}_\mu = -\frac{N_c g}{2m_v^2} B_\mu$,
%\label{V_singlet}\end{equation}
where $B_\mu =
(1/{3\pi^2})\varepsilon_{\mu\nu\lambda\rho}\mbox{Tr}
          (a^\nu a^\lambda a^\rho)$
and $v_\mu(a_\mu)=\frac{1}{2}(L_\mu +(-) R_\mu)$. If the vector
mesons are integrated out by using these constraints, the
Lagrangian (\ref{Lag})+(\ref{Lag_an}) reduces to the Lagrangian
for the pseudoscalars with the quartic Skyrme term and a special
sixth order derivative term. All the interactions between the
pseudoscalars mediated by the vector mesons then become contact
terms in the heavy mass limit.

The first step of the bound state approach is to find the static
$B=1$ soliton solutions $U_{(0)}(\vec{r})$ and
$V_{(0)}^\mu(\vec{r})$ in the non-strangeness sector. Next, the
fluctuations on top of the classical soliton configuration in the
strangeness direction -- corresponding to ${{\cal O}} (1)$ -- can
be incorporated through the Ansatz\cite{SMNR89}
\begin{equation}
\xi^\dagger_L = \xi_0 \sqrt{U_K}, \hskip 2em
\xi_R = \sqrt{U_K} \xi_0,
\end{equation}
so that $\xi_L^\dagger \xi_R$ becomes the Callan-Klebanov
Ansatz $\xi_0 U_K \xi_0$ and
\begin{equation}
V_\mu = V^{(0)}_\mu + \frac{g}{2}
\left(  \begin{array}{cc}
0 & \sqrt2 K^*_\mu \\
\sqrt2 K^{*\dagger}_\mu & 0
\end{array} \right).
\end{equation}
Substituting these Ansatz into the Lagrangian and keeping the
terms to second order in the fluctuating fields, we obtain
\begin{eqnarray}
L&=&
(D^{(0)}_\mu K)^\dagger D^{(0)\mu} K
- m_K^2 K^\dagger K
\nonumber \\
& &
\hskip -1em
+ \textstyle\frac12 m_\pi^2 (1-\cos F) K^\dagger K
+ K^\dagger a^{(0)}_\mu a^{\mu(0)} K
\nonumber \\
& &
\hskip -1em
- \frac{a}{2}  \mbox{Tr} \big[
( v^{(0)}_\mu + i q^{(0)}_\mu )
( K(D^{(0)}_\mu K)^\dagger - D^{(0)}_\mu K K^\dagger ) \big]
\nonumber \\
& &
\hskip -1em
 +a f^2_\pi g^2
\left( \frac{1}{gf_\pi} a^{(0)}_\mu K + K^*_\mu \right)^\dagger
\left( \frac{1}{gf_\pi} a^{\mu(0)} K + K^{* \mu} \right)
\nonumber \\
& &
\hskip -1em
- \frac{1}{2} ( K^{* \dagger}_{\mu\nu} K^{* \mu \nu}
               + 2i K^{* \dagger}_\mu q^{\mu\nu (0)} K^*_\nu )
\label{K-lag}\\ %--- equation number here ---
& &
\hskip -1em
 +\left( \frac{iN_c}{4f^2_\pi} \right)
\left\{
  (\textstyle \frac{2}{2} - \frac{3}{2} )
  B^{\mu(0)} \left( (D^{(0)}_\mu K)^\dagger K -K^\dagger D^{(0)}_\mu K \right)
\right.
\nonumber \\
& &
\hskip 5em
  + K^\dagger (i\frac{3g}{2}\omega_0 B^{(0)}_0) K
\nonumber \\
& &
\hskip -1em
  - \frac{2gf_\pi}{3\pi^2} \epsilon^{\mu \nu \lambda \rho}
  \big(
     K^*_\mu a^{(0)}_\nu a^{(0)}_\lambda D^{(0)}_\rho K
     - (D^{(0)}_\mu K)^\dagger a^{(0)}_\nu a^{(0)}_\lambda K^*_\rho
  \big)
\nonumber \\
& &
\hskip -1em
\left.
    - \frac{2}{3\pi^2} \epsilon^{\mu \nu \lambda \rho}
    (D^{(0)}_\mu K)^\dagger
  \big\{
      a^{(0)}_\nu ,
      (v^{(0)}_\lambda + iq^{(0)}_\lambda)
  \big\}_{+}
     D^{(0)}_\rho K
\right\},
\nonumber
\end{eqnarray}
where $K^*_{\mu\nu} = {\cal D}^{(0)}_\mu K^*_\nu - {\cal
D}^{(0)}_\nu K^*_\mu$, and $q_{\mu\nu}^{(0)} = {\cal D}^{(0)}_\mu
q_\nu^{(0)} - {\cal D}^{(0)}_\nu q_\mu^{(0)}$. Note that we have
two kinds of covariant derivatives, one with the induced vector
fields $v_\mu^{(0)}$, $D^{(0)}_\mu = (\partial_\mu + v_\mu^{(0)}
)$, and the other with the vector meson background $q^{(0)}_\mu$,
${\cal D}^{(0)}_\mu = (\partial_\mu - i q^{(0)}_\mu)$.

\begin{figure}[tb]
\centerline{\epsfig{file=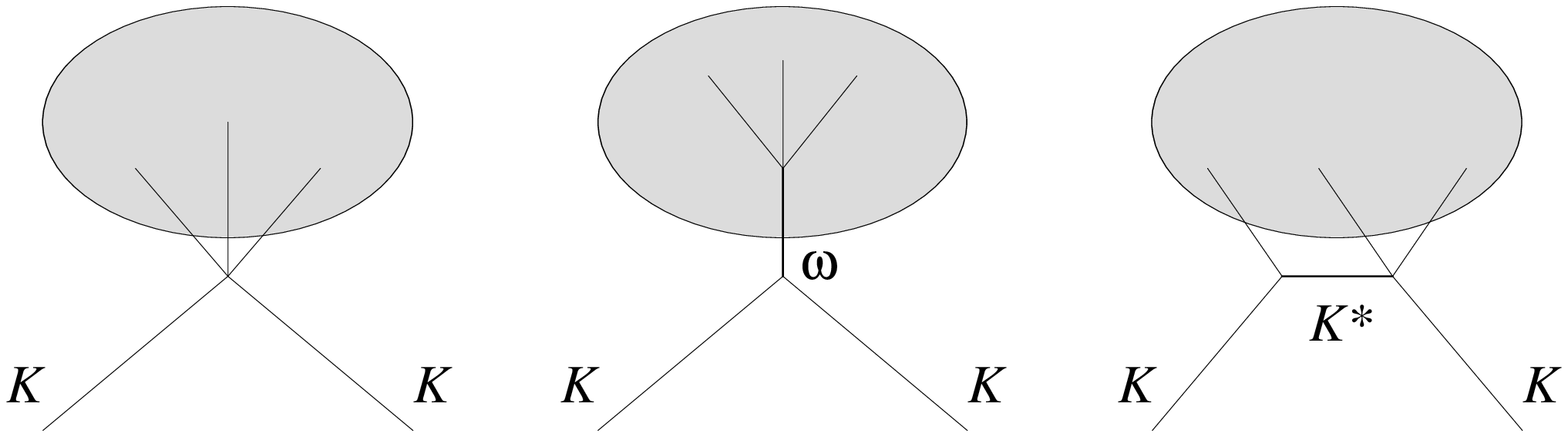,width=1.5cm,height=6.5cm,angle=270}}
\setlength{\unitlength}{1mm}
\begin{picture}(80,8)
\put(15,4){\makebox(0,0)[c]{$-\frac32$}}
\put(40,4){\makebox(0,0)[c]{$\frac12\displaystyle
          \frac{\omega_{(0)}}{\omega^\infty_{(0)}}$}}
\put(65,4){\makebox(0,0)[c]{$\frac22\displaystyle
          \frac{K^*_\mu}{K^{*\infty}_\mu}$}}
\end{picture}

\caption{Schematic illustration of the WZ-like terms due to vector
mesons in the Lagrangian (\ref{K-lag}). The baryon number density
is given by the three pion lines stemming from the shaded area,
i.e., the soliton. In the limit of infinitely heavy meson masses,
an exact cancellation takes place among the three terms.
\label{WZ:illust}}
\end{figure}

The key point of our approach is that the presence of $K^*$
modifies especially the terms of first order in time derivative,
which distinguishes $S=\pm1$ fluctuations. It should be recalled
that in the model with pseudoscalars only as in \cite{CK} such a
term arises uniquely from the ``irreducible" Wess-Zumino term. We
shall refer to the terms that distinguish $S=\pm 1$ fluctuations
{\it other} than the irreducible WZ term as ``WZ-like terms." In
Eq.(\ref{K-lag}), we have many such WZ-like terms. They originate
not only from the homogeneous Wess-Zumino term but also from the
covariant derivatives with the static $\omega$ configuration as a
gauge potential. The most striking result is that in addition to
the irreducible WZ term, $+({iN_c}/{4f^2_\pi}) B^{\mu(0)} (
(D^{(0)}_\mu K)^\dagger K -K^\dagger D^{(0)}_\mu K ) \times
\frac22$, of the Callan-Klebanov Lagrangian, there is an
additional term with a factor $-\frac{3}{2}$ from the homogeneous
part of the anomalous Lagrangian (\ref{Lag_an}). However, this
does not necessarily mean that the $S=+1$ fluctuation will now
feel an attractive interaction with the skyrmion. One can easily
check that when the vector mesons are integrated out, the
additional terms are exactly cancelled by (i) the term with the
non-vanishing time component of $q_{\mu=0}^{(0)}$(=the classical
$\omega$ configuration) in the third line of eq.(\ref{K-lag}),
with the $\omega_{(0)}$ replaced by its infinite mass limit
$\omega^\infty_{(0)}$,
%as eq.(\ref{V_singlet}),
and (ii) the eighth line in eq.(\ref{K-lag}), with $K^*_\mu$'s
replaced by their infinite mass limit which can be read off from
the fourth line as
\begin{equation}
K^{*\infty}_\mu = \frac{-1}{gf_\pi}a_\mu^{(0)} K.
\label{K*}\end{equation} The sources of the various terms that
cancel in the infinite mass limit are schematically illustrated in
Fig.~\ref{WZ:illust}.

%%%%%%%%%%%%%%%%%%%%%%%%%%%%%% FIGURE %%%%%%%%%%%%%%%%%%%%%%%%%%%%%%%%%%%%%%
%%%%%%%%%%%%%%%%%%%%%%%%%%%%%%%%%%%%%%%%%%%%%%%%%%%%%%%%%%%%%%%%%%%%%%%%%

However, with finite vector meson masses as required in HLS
theory, the cancellation becomes imperfect. The deviation of
$\omega_{(0)}$ from its infinite mass limit $\omega^\infty_{(0)}$
is such that $\langle \omega_{(0)}/\omega^\infty_{(0)}\rangle < 1$
independently of fluctuations in the strangeness direction. Thus,
the incomplete cancellation effectively reduces the net strength
of the Wess-Zumino attraction in the $S=-1$ channel, which was the
motivation of introducing the vectors in Ref.\cite{SMNR89} to
solve the over-binding problem for the hyperons in the
Callan-Klebanov model. What was not noticed in \cite{SMNR89} was
however that the $K^*_\mu$ solution depends strongly on its
strangeness. In the case of the $S=-1$ fluctuation, the closer it
is to $K^{*\infty}_\mu$, the more exact the cancelation among the
WZ-like terms becomes. Thus, the net strength of the Wess-Zumino
attraction becomes stronger approaching the irreducible strength.
As for the $S=+1$ fluctuation, however, the opposite phenomenon
takes place; the more $K^*_\mu$ deviates from its infinite mass
limit, the less cancellation among the WZ-like terms occurs. The
sum of the irreducible Wess-Zumino term and the WZ-like terms can
then give an attraction although the fourth term in the Lagrangian
prevents $K^*_\mu$ from deviating too much from $K^{*\infty}_\mu$.

Now, our task is to solve the equations of motion for $K$ and
$K^*$ moving in the background potentials provided by the static
soliton configuration sitting at the origin. This is
straightforward though tedious. Through the background potentials,
$K$ and $K^*$ are strongly coupled. Since the soliton solution is
invariant only under the simultaneous rotations in the spatial and
isospin spaces, the eigenstates are classified by their ``grand
spin" quantum number $\lambda$ associated with the operator
$\mathbf{\Lambda} = \mathbf{L} + \mathbf{S} + \mathbf{I}$,
where $\mathbf{L}$, $\mathbf{S}$ and $\mathbf{I}$ are the angular
momentum, spin and isospin operators.

To have an initial idea, we first take the $bare$ Lagrangian
obtained by Harada and Yamawaki~\cite{HY:PR} by matching the EFT
correlators to the QCD ones at $\Lambda=1.1$ GeV. The resulting
parameters take the values: $f_\pi\approx 145$ MeV, $g\approx
3.69$ and $a\approx 1.33$. These parameters reflect in some sense
a ``large" $N_c$ limit and hence differ from the physical values
by the loop corrections down by $1/N_c$. We expect the soliton
mass to be too big compared with the physical mass of the nucleon
without ${\cal O}(N_c^0)$ (e.g., Casimir) corrections (which of
course should be calculated) but what is relevant for us is the
effective mass of $K^+$ in the background of the skyrmion. The
calculation shows that the $K^+$ is indeed bound. The binding
energy turns out to be not appreciably big, around 3 MeV, but the
sign of the mass shift seems robust.

Since nature seems to indicate that $a$ is close to 1, it would be
interesting to be able to ``dial" $a$ toward 1 and see how the
system evolves. This operation is rather intricate and
non-trivial, however, because of various ``consistency conditions"
associated with the VM as can be seen in \cite{HY:PR}. For
instance, much of the low-energy hadron dynamics can be understood
with $a$ set to its fixed-point value 1 while the other parameters
of the Lagrangian such as $g$ and $f_\pi$ depart from their fixed
point values. This comes about because there are subtle
connections between various parameters controlled by the fact that
the theory flows to the VM fixed point.

We have not done this ``self-consistent" calculation yet. Since
what we are mainly interested in is whether or not the
$K^+$-soliton system {\it can be} bound and under what conditions,
we shall simply fix all the model parameters to the empirical
values except for $a$ which we vary: $N_c=3$, $f_\pi=93$ MeV,
$m_\pi=140$ MeV, $m_K=495$ MeV and $g=5.85$. In interpreting our
results, we should keep in mind that we do have at our disposal at
least two constraints. One is that the (leading $N_c$) $bare$
Lagrangian of Harada and Yamawaki mentioned above does give a
bound $K^+$-soliton system. The other is that at $a=1$, chiral
symmetry is realized with $f_\pi=f_\sigma$ where $\la
0|A^\mu|\pi\ra=ip^\mu f_\pi$ and $\la 0|V^\mu|\sigma\ra=ip^\mu
f_\sigma$, so it is likely that the coupling of the $K^+$-soliton
bound state to the $KN$ continuum is zero in the spirit of the
arguments presented in \cite{ioffe,stech}.

\begin{figure}[t]
\centerline{\epsfig{file=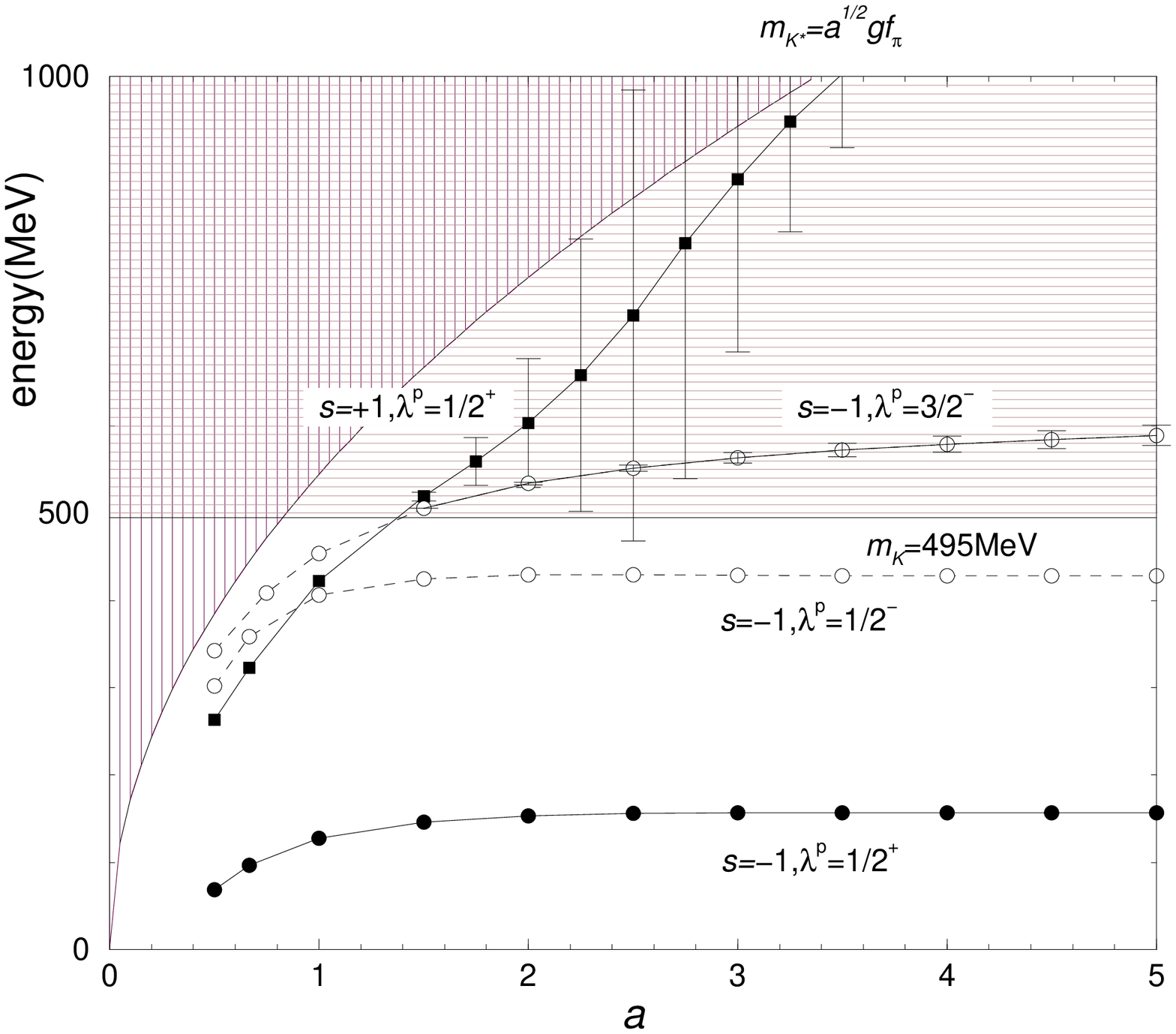,width=6.0cm,height=6.5cm,angle=270}}
\caption{The eigenenegies of $S=\pm 1$,
$\lambda^P=\frac12^\pm,\frac32^-$ states obtained for various $a$
values. The width of the resonance in the $K^+$-skyrmion channel
is given by the error bar. \label{e-vs-a}}
\end{figure}

The solutions for the $K^+$-soliton system we have so obtained are
summarized in Fig.~\ref{e-vs-a}. Plotted there are the
eigenenergies vs. $a$ of the bound states or the resonance states
found in $S=\pm 1$, $\lambda^P=\frac12^\pm,\frac32^-$ channels.
The widths of the resonance states in the kaon-soliton channel are
indicated by error bars, which however may not be interpreted in
the present form as the physical width of the $\Theta^+$. Among
others, the system should be (collective) quantized for such an
interpretation. We see in Fig.\ref{e-vs-a} that the $S=-1$ states
depend little on $a$ for the relevant range $a\geq 1$ (or as long
as the corresponding mass parameter $m_{K^*}$ is larger than
$m_K$). This means that the structure of the $S<0$ states, well
described with~\cite{SMNR89} or without~\cite{CK} vector mesons,
will not be affected by the change in $a$. There are two bound
states stable against the change of $a$ which correspond, when
quantized, to the normal $S=-1,-2$ hyperons with positive parity
or $\Lambda(1405)$ with negative parity and one bound state or
narrow-width resonance corresponding to $\Lambda(1520)$ with
negative parity. On the other hand, the $S=+1$ state is
$extremely$ sensitive to the value of $a$. Around $a=2$, the
$S=+1$, $\lambda^P=\frac12^+$ state could have a resonance state.
The energy of the resonance state is above the kaon mass but below
the mass of $K^*$. Except for those near $m_K$ threshold, the
resonance states have much too large a width to be a candidate for
the $\Theta^+$. If $a$ has the value around 1.3 as appropriate in
the large $N_c$ limit~\cite{HY:PR} where $m_{K^*}$ is comparable
to $m_K$, there can be a bound state or a sharp resonance with
very narrow width in the $S=+1$, $\lambda^P=\frac12^+$ channel.

It is noteworthy that there is no low-lying $S=+1$,
$\lambda^p=\frac32^+$ state in the model.

%------------------------------ fig 2 --------------------------------------
%----------------------------- end fig 1 -----------------------------------
Similar results are obtained for $any$ values of $m_K$; that is,
there are always stable $S=-1$ bound states below $m_K$ but for
the $S=+1$ states the bound state or a resonance state with narrow
width is possible only when there is a $K^*$ with $m_{K^*}$ close
to or even less than $m_K$. This means that to have a bound state
or a narrow resonance, a substantial modification of the vector
meson mass in the presence of the soliton is required. Whether or
not this can actually take place is not clear because of the
``self-consistency" issue mentioned above. {\it Since $m_{K^*}$ is
non-zero except at the VM, there will be no bound or narrow-width
pentaquark state in the chiral limit and hence our approach
anchored on HLS/VM does not go over to the rigid-rotor picture
even in that limit.}

In summary, when the vector mesons are incorporated \`a la
Harada-Yamawaki HLS with the VM, a bound state or a sharp
resonance can arise in the Callan-Klebanov picture of pentaquark
baryons for the range of values for $a$ implied by a variety of
considerations, $1 \lsim  a\lsim 1.4$. When quantized, it has the
quantum numbers $I=0$, $J^P=\frac12^+$ of the $\Theta^+$
pentaquark in question. Here, the vector mesons play a very
important role through a simple mechanism of level repulsion which
softens the WZ terms. We have achieved this result without
affecting the successful description of the $S<0$
hyperons~\cite{CK,SMNR89}.

%----------------------------------------------------------------------
Now the question is what our model can say about the chiral
soliton structure of the putative $\Theta^+$ believed to be seen
in the kaon-nucleon channel. Within the given scheme, the binding
appears to be quite robust as long as the value of $a$ is near 1:
It depends little on other parameters of the theory as long as
they are reasonable for non-exotic baryons. However whether the
system is bound or not is extremely sensitive to the value of $a$.
Suppose that $a$ is such that the system is bound. When the
skyrmion is collective-quantized so that both $\Theta^+$ and
nucleon have the proper quantum numbers, the bound $K^+$-soliton
system will turn into a bound state lying below the $KN$
threshold. This is clearly $not$ the $\Theta^+$ resonance one is
talking about. To the best of our knowledge, such a bound state
has not been seen in experiments. This however does not
necessarily mean that it does not exist. It could be that the
coupling to the kaon-nucleon continuum is much too weak to produce
such a state. On the other hand, if the bound $K^+$-soliton
complex could acquire additional mass by some -- so far unknown --
repulsion mechanism, such that its mass lies above the $KN$
threshold, it then could produce a narrow-width Feshbach-type
resonance discussed by Jaffe and Jain~\cite{JJ}. One could think
of the bound $K^+$-soliton as an analog of the bound
diquark-antiquark state $[ud]^2\bar{s}$ viewed as a CDD pole
discussed in \cite{JJ} with the Pauli-blocking repulsion
hypothesized in \cite{JW} playing the role of the ``unknown
mechanism" in our picture. At present, we have no idea how this
repulsion can be implemented in our model.

A more plausible possibility is that $a$ lies near 1.4 at which a
near threshold resonance is formed. The width for that resonance
will be dictated by the ${\cal O} (N_c^0)$ equation of motion with
higher order corrections strongly suppressed, so can be tiny as
indicated in Fig.2. To check whether or not this description is
viable will require the collective quantization. Even if this can
provide a description of the system, it will however be
unsatisfactory unless one can understand why $a$ is ``fine-tuned"
to a particular value. The answer may lie in understanding why
nature favors $a$ near the Georgi vector limit value
$a=1$~\cite{GV}.

Finally if the notions of HLS with VM and the soliton structure
for the pentaquarks are correct, since the HLS/VM theory moves
towards the VM (vector manifestation) fixed point as matter
density increases~\cite{HKR}, in particular, as $a$ quickly
approaches 1, the pentaquark would be {\it definitely bound} in
dense medium if it were only a resonance in free space.

The spectroscopy of the pentaquark multiplets is being worked out
and will be a subject of a forthcoming report, together with
details left out in this short article.

We are grateful for useful discussions with M. Harada, M.A. Nowak
and K. Yamawaki and for enlightening comments from I.R. Klebanov.
One of us (BYP) is grateful for the hospitality of CSSM
%(Center for the Subatomic Structure of Matter)
at the University of Adelaide,
where part of this work has been done. This work was partially
supported by KOSEF Grant R01-1999-000-00017-0~(BYP,DPM).

\end{document}